# Calculation of the dynamics of the initiation of streamer flashes that provide the NBE VHF signal profile and the VHF phase wave propagation velocity

**Alexander Yu. Kostinskiy[1], Andrei Vlasov[1] and Mikhail Fridman[1]**

[1]National Research University Higher School of Economics, Moscow, Russia, kostinsky@gmail.com

In this supplementary article to Kostinkskiy et al. (2020), we evaluate how it is possible to initiate and synchronize the start of a large number of streamer flashes, which can provide a powerful VHF signal, in the time range of ~1-3 µs. As described in Kostinskiy et al. (2020), we will assume streamer flashes occur due to the voluminous network of "air electrodes" ($E_{th}$-volumes), the number of which is dynamically supported in highly turbulent regions of a thundercloud until an extensive air shower (EAS) passes through this region. The first numerical estimates are given herein. In the near future we plan a separate article based on these estimates, where we will present the main points in more detail.

If air electrodes with an electric field $E \geq 3\, MV\, m^{-1} atm^{-1}$ (in the volume of a few centimeters in diameter) exist due to strong turbulence, statistical movement of hydrometeors and/or amplification of the electric field on hydrometeors, then any energetic charged particle passing through this volume or a photon (capable of pair production) absorbed in the volume of the electrode should lead to ordinary electron avalanches in the plasma, which will reduce the electric field inside the electrode. We estimate that for a powerful Narrow Bipolar Event (NBE) it is necessary to simultaneously accumulate about $10^7$ $km^{-3}$ of such electrodes (at a background level of cosmic rays with energies of $10^4$-$10^{11}$ eV; Sato (2015)). When $10^7$ $km^{-3}$ air electrodes exist, secondary EAS electrons having primary particle energy of $\varepsilon_0 > 10^{15}$ eV (the number of which exponentially increases in a strong electric field) can effectively synchronize the start of the phase wave of "ignition" of ordinary streamer flashes.

### Calculation of the dynamics of the formation and extinction of air electrodes depending on the height above sea level before the appearance of EAS

To calculate the background cosmic ray flux through the air electrode, we used the EXPACS (EXcel-based Program for calculating Atmospheric Cosmic-ray Spectrum) program, Sato (2015), which allows us to accurately estimate the cosmic ray flux at an altitude of 0-62 km anywhere on the Earth, for day, month, and year.

As an example, we evaluated the cosmic ray flux for a specific NBE at 00:04:38 UT on 30 July 2016 (Bandara et al., 2019), which occurred at an altitude of 6.1 km, at the latitude and longitude of the University of Mississippi (Oxford, MS, USA). We found the diameter of the air electrode based on the Meek's criterion (Raizer, 1991) for the electric field $E \geq 3\, MV\, m^{-1} atm^{-1}$ at this height and it was about 5 cm. Electrons play the main role in the background ionization of the atmosphere at these altitudes. It is also necessary to consider the contribution of positrons and photons, Sato (2015). The EXPACS program calculated that 1.34 electrons, 0.0565 positrons and 205 photons pass in one centimeter per second at a height of 6.1 km. Each electron and positron flying through the air electrode leads to the initiation of electron avalanches, since on each centimeter the paths generate at this height about 35 secondary electrons, which leads to guaranteed initiation of avalanches. High-energy photons have weak absorption in air and all 205 photons produce only 0.13 absorption per second per square centimeter of cross-section with an air electrode diameter of 5 cm. Thus, the flux of all ionizing particles will be equal to 1.53 $cm^{-2}$ $s^{-1}$. Consequently, an air electrode with a diameter of 5 cm will be ionized at an altitude of 6 km on average with a frequency of $\nu_{ae} \approx 30\ s^{-1}$.

To estimate the dynamics of the formation and extinction of air electrodes, we compose an equation for the number of air electrodes in a certain cloud volume,

$$\frac{dN_{ae}}{dt} = a - \nu_{ae} \cdot N_{ae} \ , \qquad \text{(s1)}$$

$N_{ae}$ is the number density of air electrodes [$L^{-3}$], $t$ is the time [S], $a$ [$L^{-3}S^{-1}$] is the rate of formation of air electrodes due to turbulence, statistical fluctuations of the electric field, and amplification of the electric field by hydrometeors, $\nu_{ae}$ is the frequency of extinction of air electrodes [$S^{-1}$]. In a first approximation, we consider the rates of formation and extinction of air electrodes to be constant.

The solution to this equation will be

$$N_{ae} = N_{ae}^{0} e^{-\nu_{ae} t} + \frac{a}{\nu_{ae}}\left(1 - e^{-\nu_{ae} t}\right) \qquad \text{(s2)}$$

$N_{ae}^{0}$ is the number density of air electrodes at the arrival time of the EAS. In ~150 ms, this equation reaches the stationary solution $N_{ae} \approx \frac{a}{\nu_{ae}}$.

Kostinskiy et al. (2020, section 5.1.2) estimated that in a cubic kilometer ($10^9$ $m^3$) we need approximately $10^7$ air electrodes for a strong NBE, that is, we need one air electrode per 100 $m^3$. For an average of at least one air electrode to be in this volume, the extinction should be balanced by the formation of the electrodes, i.e. $\frac{a}{\nu_{ae}} \approx 1$.

Thus, for a given height of 6 km, the formation frequency should be at least 30 air electrodes per second in a volume of 100 $m^3$. Therefore, to ensure a strong NBE, one electrode should form (on average) once a second in a volume of 1.5x1.5x1.5 $m^3$ and it should exist (on average) for about 33 ms until the ionizing background cosmic particle initiates avalanches in it.

For a height of 9 km, the diameter of the air electrode according to Meek's criterion will increase to ~7 cm (with an electric field $E \geq 3\ MV\ m^{-1} atm^{-1}$), and the ionization frequency according to similar calculations using EXPACS will increase in three times to $\nu_{cm^2} \approx 5\ \frac{1}{cm^2 s}$. Therefore, the lifetime of the air electrode will be reduced to 5.2 ms, and the side of the cube of air where the air electrode should appear will become about 0.8 m long.

If we take the measurement results for 226 NBEs on one storm day in Florida by Karunarathne et al. (2014) then at the average height for NBE formation of 13 km, the ionization frequency reaches $\nu_{cm^2} = 10.6\ \frac{1}{cm^2 s}$. For a height of 13 km, the diameter of the air electrode according to Meek's criterion will increase to 11.2 cm (with $E \approx 3\ MV\ m^{-1} atm^{-1}$). The ionization frequency becomes equal to $\nu_{ae}$ = 1052 $s^{-1}$. The lifetime of the air electrode will decrease to 0.95 ms, and the side of the cube will become about 0.46 m long. At an altitude of 16 km, the ionization frequency will be $\nu_{ae}$ = 12.25 $\frac{1}{cm^2 s}$, and the diameter of the air electrode according to Meek's criterion will increase to 16.07 cm ($E \approx 3\ MV\ m^{-1} atm^{-1}$). The ionization frequency becomes $\nu_{ae}$= 2485 $s^{-1}$. The lifetime of the air electrode will decrease to 0.4 ms, and the side of the cube will become about 0.34 m long. The calculation results for the breakdown field $E \approx 3\ MV\ m^{-1} atm^{-1}$ are presented in Table S.1.

Table S.1 Calculation results for the breakdown field $E \approx 3\ MV\ m^{-1} atm^{-1}$

| Altitude [km] | $\alpha_{eff}$ [cm$^{-1}$] | Diameter $k_M[cm]$ | $\nu_{cm^2} \cdot \frac{1}{cm^2 s}$ | Ionization Frequency $\nu_{ae}, s^{-1}$ | Lifetime $\tau_{ae}, s$ |
|---|---|---|---|---|---|
| 0 | 8.36 | 2.39 | 0.0381 | 0.17 | 5.85 |
| 6 | 8.36 | 4.88 | 1.4575 | 27.26 | 0.037 |
| 9 | 8.36 | 6.98 | 5.026 | 192.32 | 0.0052 |
| 13 | 8.36 | 11.24 | 10.6 | 1051.79 | 0.000951 |
| 16 | 8.36 | 16.07 | 12.25 | 2484.61 | 0.000402 |

It should be noted that these estimates for altitudes of 9-16 km give very short lifetimes of air electrodes under ordinary cosmic ray conditions with a minimum electric field inside the air electrode ($E \approx 3\ MV\ m^{-1} atm^{-1}$). For a simple comparison, consider that if the air at 13 km altitude in a thundercloud moves at a speed of 30-40 m/s, then it will pass through only 1 - 5 cm in 0.2 - 1 ms. Thus, the short predicted lifetime of air electrodes at typically high formation heights of strong NBEs requires a more careful analysis.

There are two possible ways the above analysis may differ for high altitudes. The first possibility is that processes at high altitudes are so intense that they provide a very high rate of production of air electrodes. The second possibility is that a real air electrode has smaller diameter than Meek's criterion determines for a minimum breakdown field $E \approx 3\ MV\ m^{-1} atm^{-1}$. The diameter of the air electrode can be reduced if the electric field inside the volume of the air electrode exceeds $E \approx 3\ MV\ m^{-1} atm^{-1}$. This second possibility seems more realistic to us. In the range of electric fields 35.7-142 $kV\ cm^{-1} atm^{-1}$ (3.57-14.2 $MV\ m^{-1} atm^{-1}$) Townsend's ionization coefficient $\alpha_{eff}$ for air is well described by the interpolation formula (Raizer, 1991, p. 57), which we rewrite for atmospheric pressure in the form convenient for use

$$\alpha_{eff} = 8.892 \cdot 10^{-2} \cdot (1.233 \cdot E - 32.2)^2\ ;\ \alpha_{eff}[cm^{-1}];\ E\left[\frac{kV}{cm}\right] \qquad \text{(s3)}$$

Meek's criterion, depending on the altitude, changes exponentially, and the diameter of the air electrode required to start the streamer flash will also grow exponentially with height. It can be described with this expression

$$k_M[cm] \approx \frac{20}{\alpha_{eff} \cdot exp\left(-\frac{h}{8.4}\right)} \approx \frac{20 \cdot exp\left(\frac{h}{8.4}\right)}{\alpha_{eff}}. \qquad \text{(s4)}$$

The frequency and average ionization time of the air electrode depending on the height will vary in proportion to the exponential squared and inversely to the fourth power of the electric field

$$\nu_{ae} = \nu_{cm^2} \cdot \frac{\pi (k_M)^2}{4} = \nu_{cm^2} \cdot \frac{3.97 \cdot 10^4 \cdot \left(exp\left(\frac{h}{8.4}\right)\right)^2}{(1.233 \cdot E - 32.2)^4}\ ;\ E\left[\frac{kV}{cm}\right], h[km],\quad \tau_{ae} = \frac{1}{\nu_{ae}}. \qquad \text{(s5)}$$

Based on these formulas, we can find the electric fields inside the air electrodes that provide reasonable electrode lifetimes of $\tau_{ae} \geq 30$ ms at high altitudes. Table S.2 lists various results of these calculations.

Considering NBEs in Florida conditions (e.g., Karunarathne et al., 2014), and requiring that the lifetime of the air electrode remains within 30-50 ms, then the electric field in the volume of the air electrode should be 1.33 $MV\ m^{-1}$ (4 $MV\ m^{-1} atm^{-1}$) at an altitude of 9 km; 950 $kV\ m^{-1}$ (4,5 $MV\ m^{-1} atm^{-1}$) at 13 km altitude; and 740 $kV\ m^{-1}$ (5 $MV\ m^{-1} atm^{-1}$) at 16 km altitude (Table S.2).

Table S.2 Calculation results for the breakdown field $E \approx 3 - 5\ MV\ m^{-1}atm^{-1}$

| Altitude km | E [kV/(cm atm)] | E [MV/(m atm)] | $\alpha_{eff}$ [cm$^{-1}$] | Diameter $k_M[cm]$ | $\nu_{cm^2}[\frac{1}{cm^2 s}]$ | Ionization frequency $\nu_{ae}\ [s^{-1}]$ | Lifetime $\tau_{ae}[s]$ |
|---|---|---|---|---|---|---|---|
| 0 | 30 | 3 | 8.36 | 2.39 | 0.038 | 0.17 | 5.85 |
| 6 | 30 | 3 | 8.36 | 4.88 | 1.46 | 27.26 | 0.037 |
| 9 | 40 | 4 | 26.06 | 2.24 | 5.03 | 19.80 | 0.050 |
| 13 | 45 | 4,5 | 48.31 | 1.95 | 10.60 | 31.66 | 0.031 |
| 16 | 50 | 5 | 77.13 | 1.742 | 12.25 | 29.20 | 0.034 |

Thus, to provide the VHF signal characteristic of NBE, the size of the air electrodes necessary for the initiation of streamers is limited from above by background cosmic radiation and varies from ~5 cm (at an altitude of 6 km) to ~1.7 cm (at an altitude of 16 km). Moreover, to ensure such a size for altitudes of 9-16 km, the electric field inside the air electrodes should be everywhere 40-50 kV $kV\ cm^{-1}atm^{-1}$ inside the entire diameter. If the air electrodes are smaller or the electric field inside the air electrodes is less than these values, but more than 30 $kV\ cm^{-1}atm^{-1}$, then electron avalanches may appear inside them but these do not turn into a streamer.

**Synchronized streamer flashes triggering with EAS-RREA - mechanism**

In the previous section, we examined the dynamics of the appearance and discharge of air electrodes. In this part, we numerically evaluate the synchronization mechanism of the start of streamer flashes due to secondary electrons (and positrons) of EASs, the number of which exponentially increases in the electric field due to relativistic runaway electron avalanche (RREA).

This calculation is a lower bound for at least two reasons. The first reason is related to the fact that in our calculation model the EAS falls vertically on the EE-volume. In the field configuration we consider (with negative charge at the top and positive charge at the bottom), the secondary EAS electrons are amplified in the electric field; the positrons (which are only several times smaller than electrons, Grieder (2010); Rutjes et al. (2019)) are decelerated and also create runaway electrons. When the EAS is incident at angles greater than $45^0$, many EAS positrons due to Coulomb scattering will begin to move in the direction of the negative charge, increasing their energy. In this way, positrons will create new electron avalanches that multiply downward in the direction of the positive charge (a kind of analogue of the feedback mechanism of Dwyer et al. (2012)). Secondly, we take into account in the calculation the exponential multiplication of only those secondary electrons and positrons that fell to the boundary of the region of strong electric field; hereinafter these electrons and positrons will be called seed electrons. We do not take into account the fact that in the process of EAS movement through a region of a strong electric field, energetic particles create additional electrons and positrons, which would also undergo the process of exponential reproduction.

According to the Mechanism of Kostinskiy et al. (2020), a necessary condition for NBE occurrence is that within an EE-volume with a size of ~0.1-1 km$^3$ there will be a rapid production of air electrodes in every 30-50 ms period (Table S5.2) and in each volume of ~2x2x2-5x5x5 m$^3$ ($10^1$-$10^2$ m$^3$). A sufficient condition for NBE occurrence is the almost simultaneous start of streamer flashes in the EE-volume. The start of streamer flashes is initiated by energetic electrons ($\varepsilon_e$> 500 keV), which cross the air electrodes. If the Meek's criterion is fulfilled (Raizer, 1991), then avalanches turn into streamers, since when an energetic electron moves through the electrode, about 75 thermal electrons are formed on each centimeter (at atmospheric pressure), which ensures the probability of starting avalanches and streamers is close to ~1.

Gurevich et al. (1999) were the first to pay attention to the possible decisive role of EASs in the initiation of lightning, since EASs with energies in the range $10^{15}$ eV ≤ $\varepsilon_0$ ≤ $10^{17}$ eV generate a large number of secondary energetic and thermal electrons and positrons, the number of which can exponentially increase due to runaway electrons. Gurevich et al. (1999) hypothesized that using EAS secondary electrons and an electron runaway mechanism could achieve such a high electron concentration in the vicinity of the primary cosmic particle that the start of multiple classical streamers would become possible. Further numerical calculations showed that this mechanism of streamer initiation is very difficult to implement, since the electron density required to start streamers turned out to be lower by several orders of magnitude (Dwyer, 2010; Babich & Bochkov 2011; Rutjes et al., 2019).

In the proposed Mechanism (Kostinskiy et al., 2020), the role of EAS is fundamentally different from the role of EAS in the mechanism of Gurevich et al. (1999). Kostinskiy et al. (2020) assume that EAS does not initiate a streamer near the EAS axis, but tens and hundreds of thousands of streamer flashes that are located in a volume of ~0.1-1 $km^3$. The main role is played by secondary electrons (and positrons) remote from the EAS axis, distributed over a region of strong electric fields in a highly turbulent part of the cloud (EE-volume). Streamers themselves are initiated by energetic electrons in areas with a local electric field above the breakdown (air electrodes several centimeters in diameter), which are formed due to turbulent motion, hydrodynamic instabilities and, possibly, an increase in the electric field near hydrometeors.

The birth of a typical NBE requires extreme turbulence and energetic EAS entering this region of the cloud. Weaker initiating events (IE) require fewer streamer flashes. The formation of IE can occur with less turbulence or when a less energetic EAS enters the highly turbulent region. Therefore, it is likely that IEs are triggered much more often than typical NBEs.

Gurevich et al. (1999) estimated that streamers near the primary EAS particle in their model require EAS with a primary particle energy $\varepsilon_0 > 10^{15}$ eV. Rutjes et al. (2019) and Dubinova et al., (2015) also identified this EAS energy range ($10^{15}$ eV ≤ $\varepsilon_0$ ≤ $10^{17}$ eV), which can play a major role in the possible initiation of streamers on hydrometeors, including the simultaneous initiation of several streamers that will provide a fast positive breakdown (FPB) mechanism proposed in Rison et al. (2016). According to the first estimates, for the implementation of our Mechanism for synchronizing streamer flashes, primary particles from a close energy range (5x$10^{14}$ eV ≤ $\varepsilon_0$ ≤ 5x$10^{15}$ eV) can play the largest role.

The frequency of the appearance of cosmic rays at the atmospheric boundary is equal to the integral of the known experimental distribution, which changes the exponent several times across the useful range (Tanabashi M. et al., 2018; Budnev et al, 2020). For estimates, we will use an approximation of this distribution, which fully satisfies our requirements for the accuracy of estimates: $\frac{dN_{\varepsilon_0}}{dE} \approx a \cdot E^{\mu},\ km^{-2}s^{-1}sr^{-1}PeV^{-1}$ , where *a=2.66*, *μ*=—2.73 for the range $10^{14}$eV ≤ $\varepsilon_0$ ≤ 3x$10^{15}$eV and *a=4.14*, *μ*≈—3 for the range 3x$10^{15}$eV ≤ $\varepsilon_0$ ≤ $10^{17}$eV. $E$ is applied in units of PeV ($10^{15}$ eV). The integral of this distribution will be $N_{\varepsilon_0}(0.1\ PeV \le \varepsilon_0 \le 100\ PeV) \approx \frac{a}{\mu+1}\left(E_1^{\mu+1} - E_2^{\mu+1}\right),\ \ km^{-2}s^{-1}sr^{-1}$.

Particles with an energy of $10^{17}$eV ≤ $\varepsilon_0$ ≤ $10^{19}$eV enter the atmosphere too rarely ($\sim 2 \cdot 10^{-4}\ km^{-2}s^{-1}sr^{-1}$) to help explain the origin of lightning. Particles with energies $\varepsilon_0 \le 10^{13}$ eV often enter the atmosphere, but they produce too few seed electrons, which can be estimated using the approximate formula $N_e^{EAS} \sim \frac{0.67 \cdot \varepsilon_0(eV)}{10^9}$ (Tanabashi et al., 2018, p. 429).

For the energy interval $10^{14}$eV ≤ $\varepsilon_0$ ≤ $10^{15}$eV, the number of primary EAS energetic particles at the atmospheric boundary will be $N_{\varepsilon_0} \approx\ 82\ km^{-2}s^{-1}sr^{-1}$ . For the energy interval $10^{15}$eV ≤ $\varepsilon_0$ ≤ $10^{16}$eV the number of primary EAS energetic particles at the atmospheric boundary will be $N_{\varepsilon_0} \approx 1.5\ km^{-2}s^{-1}sr^{-1}$ , and for the energy interval $10^{16}$eV ≤ $\varepsilon_0$ ≤ $10^{17}$eV the number of primary EAS energetic particles at the atmospheric boundary will be $N_{\varepsilon_0} \approx 2 \cdot 10^{-2}\ km^{-2}s^{-1}sr^{-1}$. Thus, the entire energy range $10^{14}$eV $\leq \varepsilon_0 \leq 10^{17}$eV requires analysis from the point of view of the initiation of streamer flashes, since the total number of particles incident on the boundary of the cloud appears reasonable.

For preliminary estimates of the number of all secondary particles at the EAS maximum in the range of $10^{14}$eV $\leq \varepsilon_0 \leq 10^{17}$eV, Dwyer (2008) uses the formula (Gaisser, 1990)

$$N_e^{EAS} \sim 5 \cdot 10^{-2} \varepsilon_0^{1.1} \quad \text{(s6)}$$

where $\varepsilon_o$ is the total energy of the cosmic-ray primary measured in GeV. For example, $N_e^{EAS}$ is 2x$10^5$ for a primary energy of $10^{15}$ eV. It should be noted that in cosmic ray physics, experimental setups usually measure electrons and positrons with an energy of not less than 10-100 MeV, but electrons with an energy of 0.5 MeV can become runaway in air, and their number is several times larger than that calculated by formula (s6) (Rutjes et al., 2019). Therefore, a primary particle with an energy of $\sim 10^{15}$ eV can produce at the EAS maximum $N_e^{EAS} \sim 10^6$ seed electrons and positrons. An estimate by the formula $N_e^{EAS} \sim \frac{0.67 \cdot \varepsilon_0(eV)}{10^9}$ (Tanabashi et al., 2018, p.429) gives a close result.

**Lateral spatial distribution of electrons in a strong electric field ($E \geq 280\ kV\ m^{-1} atm^{-1}$)**

To evaluate the initiation of NBE (IE) due to EAS-RREA avalanches, we must take into account the lateral distribution of secondary electrons and positrons of EASs, the number of which will exponentially increase in a strong electric field due to the RREA mechanism. Secondary electrons must almost simultaneously fall into a sufficient number of existing air electrodes in order to simultaneously initiate many streamer flashes.

$$\Phi_{re}^{c}(r,z) = \frac{N_0}{4\pi\left(\frac{D_\perp}{\nu}\right)(z-z_0)} \cdot exp\left(\frac{z-z_0}{\lambda} - \frac{r^2}{4\left(\frac{D_\perp}{\nu}\right)(z-z_0)}\right)[m^{-2}] \quad \text{(s7)}$$

$$\lambda = \frac{7200\ [kV]}{\left(E - 275\left[\frac{kV}{m}\right]exp\left(-\frac{h}{8.4}\right)\right)} \quad \text{(s8)}$$

$$\frac{D_\perp}{\nu} = exp\left(\frac{h}{8.4}\right)(5.86 \cdot 10^4)E^{-1.79}[m], E\ [\mathrm{kV/m}];\ \ \nu = 0.89c; h\ [km] \quad \text{(s9)}$$

The lateral diffusion coefficient (s9) depends on the electric field strength $E$ and the concentration of air molecules. In our case, the air pressure decreases exponentially with increasing height, which means that for the same electric field, the lateral diffusion coefficient increases significantly at high altitudes. In this calculation, we are not interested in longitudinal diffusion, since it does not affect the total area of the flux of secondary electrons in the electric field. Combining s7, s8, and s9, we have

$$\Phi_{re}^{c}(r,(z-z_0)) = \frac{N_0}{4\pi\left(exp\left(\frac{h}{8.4}\right)(5.86\cdot 10^4)E^{-1.79}\right)(z-z_0)} \cdot exp\left(\frac{z-z_0}{\lambda} - \frac{r^2}{4\left(exp\left(\frac{h}{8.4}\right)(5.86\cdot 10^4)E^{-1.79}\right)(z-z_0)}\right), [m^{-2}] \quad \text{(s10)}$$

The distribution (s7, s10) of Dwyer (2010) and Babich & Bochkov (2011) is written under the assumption that all electrons fall in the region of a strong electric field at one point (approximation of the Dirac delta function). But in the developed EAS, the energy range of interest to us simultaneously moves $10^4$-$10^7$ secondary particles that are distributed perpendicular to the axis of EAS propagation over hundreds of meters (Kamata & Nishimura, 1958; Grieder, 2010). Therefore, we cannot simply substitute the total number $N_0$ of EAS particles into equation (s10). The lateral distribution of EAS particles is often presented as an approximation of Nishimura-Kamata-Greizen (NKG). Therefore, a correct model of EAS propagation in a strong electric field of a thundercloud should initiate avalanches of runaway electrons according to formula (s10) at each point of the EAS cross section. Equation (s10) should take into account the number of initial electrons $N_0$ at each point of the EAS front, and not the total number of particles of the entire EAS. At each point, the number of initial electrons $N_0$ can be expressed by the formula (s11).

The NKG approximation is used for EAS characteristic estimation (Kamata & Nishimura, 1958):

$$\rho_e(R) = \frac{N_e^{EAS}}{R_M^2} \cdot C(s) \cdot \left(\frac{R}{R_M}\right)^{s-2} \cdot \left(\frac{R}{R_M} + 1\right)^{s-4.5} \quad \text{(s11)}$$

where $\rho_e(R)$ — is the relativistic electron and positron particle density at the distance $R$ from shower axis, $N_e^{EAS}$ – total number of shower particle, $R_M = 79[m] \cdot exp\left(\frac{\text{h}}{8.4}\right)$ – Mølier radii, $s$ – shower age parameter and $C(s) = 0.366 \cdot s^2 \cdot (2.07 - s)^{1.25}$ , (Hayakawa, 1969).

Simple estimates show that, at altitudes of 6–16 km, the distribution of EAS electrons will be wide (hundreds of meters). The main mechanism for increasing the lateral distribution of EAS secondary electrons is the Coulomb scattering of electrons by the nuclei of air molecules. Coulomb scattering mainly determines the root mean square radius of the EAS $\sqrt{\langle R \rangle^2}$, which includes about half of the electrons (Murzin, 1988):

$\sqrt{\langle R \rangle^2} = 0.9 \cdot \left(\frac{E_s}{\varepsilon_{cr}} \cdot t_0\right) = 0.9 \cdot R_M^{\frac{g}{cm^2}} = 0.9 \cdot 9.5 \ \frac{g}{cm^2} = 8.55 \ \frac{g}{cm^2}$ , где $E_s = 21 \ MeV$ , $t_0 \left[\frac{g}{cm^2}\right]$ — is the radiation length, $\varepsilon_{cr}$ — is the critical electron energy at which the losses due to bremsstrahlung become equal to the losses due to ionization (for air $\varepsilon_{cr} \approx 81 \ MeV$). $R_M^{\frac{g}{cm^2}} = 9.5 \ \frac{g}{cm^2}$ Mølier radii for air in units of $\frac{g}{cm^2}$. The EAS lateral expansion radius increases exponentially with height. The root mean square radius of EAS $\sqrt{\langle R \rangle^2}$ in meters will be $\sqrt{\langle R \rangle^2}[m] \approx 71.1[m] \cdot exp\left(\frac{h}{8.4}\right)$. At 6 km altitude, the RMS radius of the EAS will be 144 m; at 9 km it is 208 m; for 13 km — 355 m; at 16 km altitude, it is 531 m.

This estimate shows that the secondary electrons of a well-developed EAS can seed a significant part of the EE-volume. When the secondary EAS electrons enter a region of strong electric field, each EAS electron with an energy greater than ≈ 500 keV will begin a relativistic runaway electron avalanche, which in turn also has a wide radial distribution (according to s10). The RREA electrons greatly increase the seeding of positive streamers in air electrodes.

**Algorithm for calculating EAS-RREA initiation of streamer flashes**

The general scheme of the process of synchronization of streamer flashes by the EAS-RREA avalanche between an upper negative and lower positive charge is shown in Figures 6 and 7 (section 5.1.3) of Kostinskiy et al. (2020).

Figure S2 illustrates the numerical calculation algorithm. EAS falls on the EE-volume parallel to the electric field and perpendicular to the plane ($X$-$Y$, $z= z_0(O_r)$) and moves from the point $z_0(O_r)$ along the $z$ axis. The direction of the electric field is opposite to the direction of EAS movement. The calculation is axisymmetric with respect to the z axis, which is the EAS axis. The $R$-radius is the distance to the EAS axis in the NKG approximation (s11). $R$ changes at each integration step from 5 m to the final $R_F$ value (for reasons described below we ignore contributions for R < 5 m). This circle with radius $R$ and width $dR$ is chosen so that it has the same number of seed electrons in each point and is located in the plane ($X$-$Y$, $z$=$z_0(O_r)$) of the EAS penetration into the EE-volume. From these points of the circle $R$, the electrons propagate in accordance with equation (s10) to each point $y_a(z)$. The point $y_a(z)$ changes on each integration step ($z$- $zo+dz$) from 5 m to a final distance $y_{aF}(z)$. The first 5 meters of the EAS radius require a separate description due to strong fluctuations of energetic particles (Murzin, 1988; Grieder, 2010); in this numerical approximation we neglect them without a large loss of accuracy of the overall estimate due to the small the area of this segment and its small contribution. The points $y_a(z)$>10 m from the EAS axis make the main contribution to the number of electrons that ionize the air electrodes.

Due to the axial symmetry of the problem, it suffices to obtain the value of the electron flux along any radius running perpendicularly from the $z$ axis to the point $y_{aF}(z)$. We can express this value by the distribution $N_{y_a}(z) = f(y_a(z))$, and obtain it by changing the position of the point $y_a(z)$ (red arrow in Figure S2).

For a given $R$ we determine the density of seed electrons $\rho_e(R)$ due to the NKG - equation (s11) for $s=0.9$, and we obtain a circle centered at the point $O_r(z_0)$ with a constant density of seed electrons at each point of the circle. This circle is located in the plane where the EAS penetrates the region of a thundercloud with a strong electric field E (at the boundary of a strong electric field). Above height $z_0$ we assume in the calculation that the electric field was zero. The value of the EAS age $s=0.9$ corresponds in this energy range $\varepsilon_0$ (of the primary particle) to the EAS developed in space and time with $10^5$-$10^7$ secondary electrons (Grieder, 2010; Rutjes et al., 2019). We call an electric field strong if it can support not only runaway electrons $E > 284$ $kV/(m{\cdot}atm)$, but can also support the movement of positive streamers $E > 450$ $kV/(m{\cdot}atm)$. A circle of radius $R$ is the "supplier" of electrons $N_r$ to the point $y_a(z)$ in accordance with the distribution below (s12), which is obtained by substituting in (s10) the distribution NKG (s11). In the formula (s10), $r$ is the distance in the $X$-$Y$ plane from an arbitrary point of the circle $R$ to the point $y_a(z)$. We will find the distance $r$ by the well-known formula below (s13), where we placed the EAS axis at the center of the circle, which in turn is placed at the center of the coordinate system.

The point $y_a(z)$ changes along the y axis, which implies that $x_a = 0$. Also, when integrating, it is necessary to take into account two values $\pm\left(R^2 - x_i^2\right)^{0.5}$, since both points $(x_i, y_i^+)$, $(x_i, y_i^-)$ contribute to the electron flux $N_{y_a}$ (Figure S2). In addition, the solution for each point $(x_i, y_i^+)$, $(x_i, y_i^-)$ should be multiplied by 2, because the points of the left side of the circle R (Figure S2), due to symmetry, make exactly the same contribution to $N_{y_a}$. The sum of the contribution of all points of the circle $R$ can be obtained by changing $x_i$ from 5 to $R$. $R$ varies from 5 m to the "edge" of the EAS ($R_F$). The sum of the contributions is the electron flux of all circles $R$ and gives the total electron flux $N_{y_a}$ at the point $y_a(z)$ (in each layer $z$-$z_0+dz$).

$$N_r(z-zo) = \frac{0.361\cdot\frac{N_e^{EAS}}{R_M^2}\cdot\left(\frac{R}{R_M}\right)^{-1.1}\left(\frac{R}{R_M}+1\right)^{-3.6}}{4\pi\left(exp\left(\frac{h}{8.4}\right)(5.86\cdot 10^4)E^{-1.79}\right)(z-z_0)}\cdot exp\left(\frac{(z-z_0)}{\lambda} - \frac{x_i^2+\left(\pm(R^2-x_i^2)^{0.5}-y_a(z)\right)^2}{4\left(exp\left(\frac{h}{8.4}\right)(5.86\cdot 10^4)E^{-1.79}\right)(z-z_0)}\right) \quad \text{(s12)}$$

$$r^{\pm} = ((x_i - x_a)^2 + (y_i - y_a)^2)^{0.5} = ((x_i)^2 + (y_i - y_a)^2)^{0.5} = (x_i^2 + (\pm(R^2 - x_i^2)^{0.5} - y_a)^2)^{0.5} \quad \text{(s13)}$$

The calculated electron flux at the point $(y_a(z), z - z_0)$ of the axisymmetric radial distribution in each layer of the EE-volume will be:

$$N_{y_a}(y_a(z),(z-z_0)) = \int_{R=5}^{R=R_F}\int_{x=5}^{x=R} \frac{2\cdot 0.361\cdot\frac{N_e^{EAS}}{R_M^2}\cdot\left(\frac{R}{R_M}\right)^{-1.1}\left(\frac{R}{R_M}+1\right)^{-3.6}}{4\pi\left(exp\left(\frac{h}{8.4}\right)(5.86\cdot 10^4)E^{-1.79}\right)(z-z_0)}\cdot exp\left(\frac{(z-z_0)}{\lambda} - \frac{x^2+\left(\pm(R^2-x^2)^{0.5}-y_a(z)\right)^2}{4\left(exp\left(\frac{h}{8.4}\right)(5.86\cdot 10^4)E^{-1.79}\right)(z-z_0)}\right) dxdR\,, \quad \text{(s14)}$$

After penetrating into the region of a strong electric field, the number of electrons in the avalanche of runaway electrons increases exponentially $N_e \sim N_e^{EAS}\exp(\frac{(z-z_0)}{\lambda})$ (s14), where the step of the avalanche of runaway electrons $\lambda$ is determined by the formula (s8).

Now we can estimate the number of streamer flashes initiated by this electron flux taking into account the probabilistic nature of the initiation processes. The cross section of each air electrode in a thundercloud, in accordance with previous estimates, is in the range of $10^{-2}$-$10^{-3}$ $m^{-2}$. The probability of one random electron entering the air electrode will be $p = 10^{-2}$-$10^{-3}$ (the probability is equal to the ratio of the area of the air electrode to one square meter). The probability of several such independent events is estimated using the well-known Bernoulli formula $P_n^k = C_n^k p^k (1-p)^{n-k}$. In order for the streamer flash to be initiated, it is sufficient that the cross section of the air electrode is crossed by at least one energetic electron. The probability of such an event is calculated using the simplified Bernoulli formula and it is equal to $P_{N_{y_a}} = 1 - ((1-p))^{N_{y_a}}$, where $N_{y_a}$ is the flux of energetic electrons (according to formula (s9)). The probability of initiation of a streamer flash $P_{N_{y_a}}$ at a flow of 100 electrons per square meter for an air electrode with a cross section of $10^{-2}$ $m^{-2}$ will be 63%, and for an air electrode with a cross-section of $10^{-3}$ $m^{-2}$ the same probability of initiation will be achieved when 1000 electrons per square meter fall.

Thus, we can obtain the total number of streamer flashes in the entire EE-volume, depending on the distance (or time), by integrating the number of flashes in all layers $dz$ along the $z$ axis (Figure S2):

$$n_{fl} = \int_{z=z_0}^{z=z_F} \int_{y_a=5}^{y_a=y_{aF}} \rho_{E_{th}}(z, y_a(z)) \cdot 2 \cdot \pi \cdot y_a(z) \cdot \left(1 - (1-p)^{N_{y_a}(y_a(z),z)}\right) dy_a dz \qquad \text{(s15)}$$

In the formula (s14), $\rho_{E_{th}}(z, y_a(z))$ is the density of the number of air electrodes in a cubic meter. In these calculations, we will consider this density to be a constant value in the entire volume and equal to $\rho_{E_{th}} = 10^{-2}\ m^{-3}$, based on estimates of a powerful NBE flash with a total number of air electrodes of ~ ∼$10^6$-$10^7$ км$^{-3}$ (Kostinsky et al., 2019). Another argument in favor of this value of the density of air electrons within the framework of the proposed Mechanism (Kostinskiy et al., 2020) is that at this density, streamer flashes with an average speed of 1-2x$10^6$ m/s will cover a distance of ∼5 m in 2.5-5 µs. This will create a network of plasma channels and dramatically increase their lifetime.

**RESULTS**

We will calculate the number of streamer flashes in accordance with equations (s14, s15). As noted above, these calculations provide lower bound estimates. We will carry out estimates for two altitudes of 13 and 6 km with the configuration of the field and the EAS movement in the geometry shown in Figures 7 and 8 ((section 5.1.3) of Kostinskiy et al. (2020).

In Figure S3, we show the calculation of the number of streamer flashes, which was carried out for a height of 13 km, $N_e^{EAS} = 10^6$, $R_M$ = 328 m. The electric field in this calculation was $E$ ≈ 85 kV/m (400 kV/m·atm)) and it approximately corresponds to the minimum possible electric field that can support the movement of positive streamers in conditions of extremely low humidity. The evaluation used the probability of initiation of air electrodes equal to $p$ = 0.001, which corresponds to the diameter of the air electrode Ø≈2 cm (in accordance with table S2). This diameter is determined by Meek's criterion (s4, s5) and the time of discharge of the air electrodes by background cosmic rays, which are not EASs (table S2). The density of the number of air electrodes in these estimates was considered constant and equal to $\rho_{E_{th}} = 10^{-2} m^{-3}$ in the entire volume of EAS movement in an electric field. The volume was in the form of a cylinder with a base radius of 500 m and a variable height $z$.

If we take the first 5 steps of an avalanche of runaway electrons $z$ = 1385 m (the red line in Figure S3), then the total volume of the cylinder will be 1.09 km$^3$, and the total number of air electrodes will be about $10^7$ km$^{-3}$. The number of all streamer flashes after 5 steps of an avalanche of runaway electrons in this case will be equal to 2.4x$10^5$ (Figure S3). This means that the EAS-RREA avalanche managed to initiate flashes in only 2.2% of all air electrodes. Despite this, even if the average current of one streamer flash was only 0.1 A, then the total current of all streamers will become about 24 kA. The total movement time of the EAS-RREA avalanche at this point in time is 5.2 μs.

The growth curve for the number of streamer flashes in Figure S3 as a function of the length covered by the EAS-RREA avalanche has two distinct sections. The first section of the curve corresponds to the initial penetration of EAS into the region of a strong electric field, and it is 60-70 meters (the first 200-300 ns). A sharp increase in the number of flashes from 0 to $\sim 10^3$ occurs because the EAS already has $10^6$ electrons and they immediately begin to ionize the air electrodes. This growth will correspond to the physics of the process only in the case of an almost vertical fall of EAS on a plane wide region of a strong electric field, and for other configurations of the EE-volume region it will be less sharp. Starting from 70 meters, when the EAS-RREA avalanche "completely entered" the EE-volume, we see a nearly exponential increase in the number of flashes (and, consequently, the VHF signal) over the entire length of the avalanche, depending on the distance and time of movement, which qualitatively corresponds to experimental results (Rison et al., 2016). In general, we see that the estimate gives reasonable figures with a minimum field for the distribution of streamers and a not very large total length of the EAS-RREA avalanche.

Figure S4 shows how, in this case, the number of streamer flashes ($n_{fl}$) in each transverse layer of air 1 meter thick increases with distance $z$ (and time $t$). The nearly instantaneous increase in the number of flashes at the initial moment is also clearly visible when the EAS with the number of particles $N_e^{EAS} = 10^6$ enters the EE-volume. It is important that even with such a large initial number of electrons, EAS electrons initiate no more than 5% of all EAS-RREA avalanche streamer flashes. This suggests that it is very difficult for only secondary EAS electrons, without the amplification mechanism of RREA runaway avalanches, to initiate a large number of streamer bursts in a few microseconds. This is the case with an air electrode diameter of 2 cm, and even more so for diameters of air ele ctrodes 0.1–2 mm in size, which hypothetically may be near the largest charged hydrometeors. Therefore, in our opinion, the assessment shows that in order to initiate a significant number of streamer outbreaks in volume, it is precisely the combination of EAS and RREA that is required.

**Influence of the average electric field strength on the process of initiation of streamer flashes**.

An important role can be played by the dependence of the number of streamer flashes on the electric field strength, which directly determines both the step length of the runaway electron avalanche λ (s8) and the lateral distribution of secondary electrons (s7). We continue to calculate the number of streamer flashes initiated by the EAS-RREA avalanche for an altitude of 13 km with a vertical EAS fall. The probability of initiation by a random electron was taken as $p = 0.001(\emptyset \approx 2\ cm), \rho_{E_{th}} = 10^{-2} m^{-3}$, $R_M$=328 m, $N_e^{EAS} = 10^6$. The electric field took four values: 85 kV/m (400 kV/(m·atm)), $\lambda = 277$ m; 106 kV/m (500 kV/(m·atm)), $\lambda = 153\ m$; 127.5 kV/m (600 kV/(m·atm)), $\lambda = 106\ m$; 149 kV/m (700 kV/(m·atm)), $\lambda = 81\ m$. The calculation results are shown in Figure S5. As for an electric field of 400 kV/(m·atm) (Figure S3) the growth curves for the number of streamer flashes depending on the length covered by the EAS-RREA avalanche have two distinct sections. The first section corresponds to the initial penetration of EAS into the region of a strong electric field and is 60-70 meters. Up to the line thickness on the graph, the lines begin to separate after 35 m (about 100 ns) of the path. Starting from about 70 meters and up to 500 m, for the entire length for all electric fields, we see a nearly exponential increase in the number of flashes depending on the distance (time of the avalanche). However, starting from 500 meters, the curve for 700 kV/(m·atm) becomes more and more gentle, approaching saturation. The curve for 600 kV/(m·atm), starting from about 800 meters, and the curve for 500 kV/(m·atm), starting from about 1000 meters, similarly flatten. This transition occurs between the 6th and 7th steps of the avalanche for all curves. The dependence on the magnitude of the electric field is strong. So for a field of 500 kV/(m·atm) the number of flashes equal to $10^5$ is achieved at the first $\sim 750$ meters, and for 700 kV/(m·atm) this number is achieved in only $\sim 400$ m. Thus, estimates of the number of flashes with increasing average electric field from 400

kV/(m·atm) to 700 kV/(m·atm) make it even easier to achieve the required number of streamer flashes that can provide the VHF signal accompanying a NBE.

**The effect of the volume of the EE-volume occupied by air electrodes on the process of initiation of streamer flashes due to secondary EAS electrons.**

In the next series of our calculations, at an altitude of 13 km, we evaluate the effect on the number of streamer flashes of the size of the EE-volume of a thundercloud occupied by air electrodes, reducing the radius of the EE-volume from 500 m to 250 m and 150 m. The electric field will vary from $400 - 500 \frac{kV}{m \cdot atm}\left(85 - 106\ \frac{kV}{m}\right)$. The number of seed EAS electrons will be $N_e^{EAS} \approx 10^6$. The volume of the EE-volume will be changed using the integration limit $y_{aF}$. In the above calculations, we took $y_{aF}$ = 500 m, which set the total cylinder diameter ≈ 1 km. Without changing the EAS parameters, we will define much narrower areas occupied by air electrodes $\sim 0.2\ km^3$ ($y_{aF} = 250\ m$ , $z = 1000$) and $\sim 0.07\ km^3$ ($y_{aF} = 150\ m,\ z = 1000\ m$). Estimates show (Figure S6) that the number of streamer flashes varies only slightly, despite a change in the EE-volume by more than an order of magnitude. We will consider the differences at a considerable distance of 900 m covered by the avalanche. With an electric field of 400 kV/(m·atm) and $y_{aF}$ = 250 m, the volume occupied by air electrodes will decrease by 4 times, which will lead to a decrease in the number of streamer flashes by only 23 %.Similarly, a decrease in volume by 11 times ($y_{aF}$ = 150 m) yields a decrease in the number of flashes by only a factor of 2. With an electric field of 500 kV/(m·atm), a decrease in the volume occupied by air electrodes by 4 times ($y_{aF}$ = 250 m) will lead to a decrease in the number of streamer flashes by only 24%; if the volume is reduced by 11 times ($y_{aF}$ = 150 m), then the decrease in the number of flashes is only 2.44 times.

These estimates suggest hypotheses. The main role in the initiation of streamer flashes is played by the EAS-RREA region of an avalanche with a radius of about 250 meters, which can be extended to all estimates given herein. It may be even more interesting that the narrow layers of a thundercloud with a diameter of 300-500 m that may occur during complex air currents in or near a thunderstorm cell (Karunarathna et al., 2015; Yuter & Houze, 1995), have a chance to create powerful VHF signals produced by an NBE.

**Influence of the number of seed EAS electrons on the process of initiation of streamer flashes.**

We also need to evaluate the role of the number of seed EAS electrons in the process of initiation of streamer flashes by changing the number of seed electrons — $N_e^{EAS}$ EAS. For these estimates, we use 13 km altitude. Figure S5 shows a numerical calculation performed for four electric field values, with the number of seed electrons $N_e^{EAS} = 10^6$. Figure S7 shows the calculations that were carried out for three additional values of the number of seed EAS electrons $N_e^{EAS} = 10^3; 10^4;\ 10^5$. For each of these $N_e^{EAS}$ values, the calculation was carried out for three electric fields: E= 85 kV/m (400 kV/(m·atm)), 106 kV/m (500 kV/(m·atm)), 127.5 kV/m (600 kV/(m·atm)) (also in Figure S7 the calculations for $N_e^{EAS} = 10^6$ are shown). The results shown in Figure S7 show all $N_e^{EAS}$yield similar exponential growth dynamics of the number of streamer flashes, with the exception of the first 50-60 meters, as in all Figures S3-S6. For $N_e^{EAS} = 10^3; 10^4;\ 10^5$, the curves in the second stage deviate very little from exponential development. The number of flashes at different distances of the EAS-RREA avalanche grows nearly linearly with $N_e^{EAS}$. We can clearly see that for $N_e^{EAS} = 10^3$ and an electric field of 400 kV/(m·atm), the number of streamer flashes, even with an EAS-RREA avalanche path of 1000 m, is 54. Even with an electric field of 600 kV/(m·atm) the number of streamer flashes at $N_e^{EAS} = 10^3$ is only 7.6x$10^3$ (with a total number of air electrodes in the volume of $\sim 10^7$). In this case, the average distance between streamer flashes will be in the range of 35-45 m, which will not allow creating a network of plasma channels within reasonable times.

These results show that cosmic rays with the number of seed particles $N_e^{EAS} \leq 10^3$ will not be able to slightly discharge the EE-volume and create a noticeable NBE signal. Cosmic rays with the number of seed particles $N_e^{EAS} = 10^4$ can produce the measured VHF signal at average electric fields of 500-600 kV/(m·atm). Perhaps these can also cause IEs of the weaker sort, since the distance between the streamer flashes will be in the range of 12-15 m. Overall, the estimates indicate that the powerful VHF signal that accompanies a NBE requires a number of seed electrons greater than $10^5$ and an electric field in the range 500-600 kV/(m·atm).

**The effect of pressure on the process of initiation of streamer flashes.**

As altitude in the atmosphere decreases and atmospheric pressure increases, the step of the runaway electron avalanche (RREA) and the scattering of runaway electrons by atomic nuclei rapidly decrease. In Figure S8, it is clearly seen that the number of flashes $n_{fl}$ rapidly grows with length $z$; when reaching an avalanche of 450 meters for all electric fields in the range 500-700 kV/(m·atm) $n_{fl}$ exceeds $10^5$. Figure S8 also shows that most streamer flashes at an altitude of 6 km occur within a radius of 250 m from the EAS axis, and the influence of more distant avalanche areas begins only with 600 m (700 kV/(m·atm)), 630 m (600 kV/(m·atm)) and 750 m (500 kV/(m·atm)), when the curves begin to separate. The dependence of the number of streamer flashes on the number of seed electrons $N_e^{EAS} = 10^5 \; and \; 10^6$ (Figure S9) shows that, over a length of 800-900 meters, the number of flashes becomes similar, apparently due to exhaustion the number of air electrodes per unit volume. Also, both Figures S8 and S9 clearly show the important role of the avalanche step, since the number of flashes that were initiated by EAS with an order of magnitude smaller number of seed electrons “catches up” with the number of flashes in cases that occurred at lower electric fields. This is not surprising, since the total number of electrons due to the small step of the avalanche for large fields increases rapidly and begins to exceed the total number of electrons with a smaller electric field, but with a large number of seed electrons.

## Discussion of the first estimates

In the case of validity of Mechanism (Kostinskiy et al., 2020), estimates indicate that only an EAS-RREA avalanche with the number of seed particles $N_e^{EAS} > 10^5$ can provide the necessary number of electrons and positrons for the synchronous initiation of streamer flashes to explain the observed VHF signals that accompany powerful NBEs over their known range of altitudes. EAS provides not only a large number of seed electrons and positrons, but also their simultaneous wide initial lateral distribution. EASs ($\varepsilon_0 \leq 10^{17}$ eV) alone, without the RREA mechanism, are not able to initiate a sufficient number of electrons for synchronization. Likewise, for the RREA mechanism in realistic electric fields (400-700 kV/(m· atm)), without a sufficient number of seed electrons, EASs cannot create the required number of electrons in the volume to initiate a large number of streamer flashes. In order to reach the initial number of EAS seed electrons, the RREA mechanism in realistic electric fields needs to take $\ln(10^6) = 13.8\,\lambda$ steps;. after these, an additional 3-7 steps are needed for the RREA avalanche electrons to initiate a sufficient number of streamer flashes.

Even when evaluated with lower bound estimates, as done herein, EAS-RREA avalanches with the number of electrons $N_e^{EAS} \approx 10^5 - 10^6$ in electric fields (400-700 kV/(m·atm)) provide the necessary electron flux for simultaneous synchronization through several microseconds of many streamer flashes in the EE-volume. The EAS-RREA initiation of streamer flashes can provide many weak-IE events that require fewer secondary EAS particles and fewer streamer flashes than a strong NBE.

If the proposed Mechanism of Kostinskiy et al. (2020) is correct, then the IE energy spectrum may be semi-continuous, from strong NBE to weakest IE parameters. This hypothesis can be suggested, since the statistics of interactions between EAS and turbulent charged regions will lead to a large number of combinations between the spatial cross section of turbulent regions and EAS of various energies and ages falling in the EE-volume from different angles. Furthermore, a detailed study of NBE parameters at different heights may provide information on the most powerful turbulent flows inside a thundercloud (indicating regions that may have large and opposite charges in close proximity).

Finally, it is interesting to consider that the EE-volumes may serve as original “streamer cameras” for studying EASs at different altitudes. This idea is a continuation of the idea Gurevich et al. (1999, 2004), which previously suggested that a thundercloud be considered a “spark chamber”.

**Acknowledgements**

We are very grateful to Thomas Marshall and Maribeth Stolzenburg for a fruitful discussion of the main provisions of the article, important criticisms and assistance in preparing the manuscript for publication as well as to Gagik Hovsepyan and Ashot Chilingarian for the recommendation to use the EXPACS program.

We are grateful to Vladimir Rakov for a detailed discussion of the most controversial statements of the article, to Stanislav Davydenko, Alexander Litvak, Evgeni Mareev for discussions that showed the need to conduct more detailed numerical estimates of the synchronization in a significant volume of the cloud of many streamer flashes.

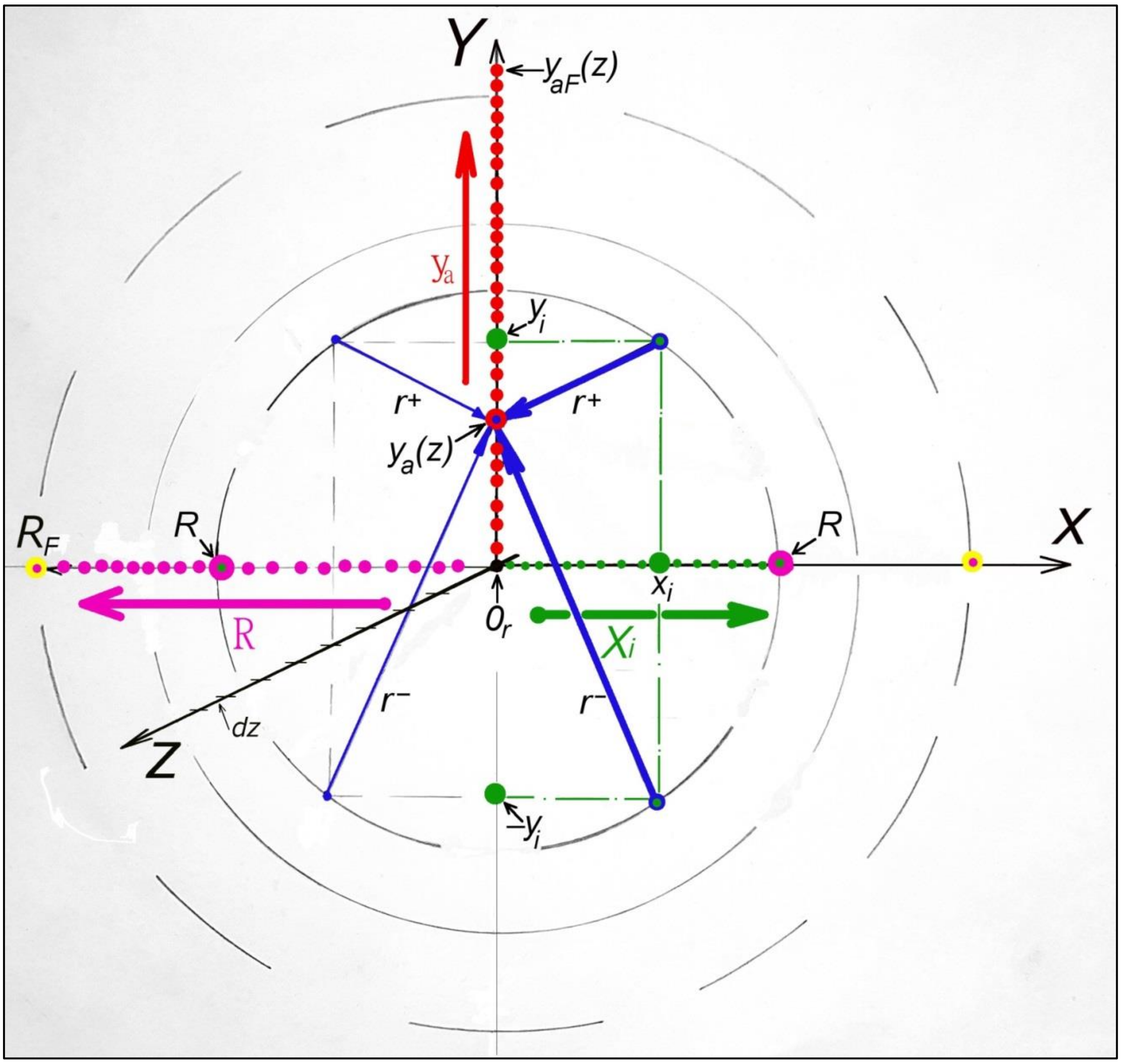

**Supplementary Figure S.2:** Scheme for calculating the flow of electrons crossing the region of a thundercloud with a strong electric field. EAS falls on the EE-volume parallel to the electric field and perpendicular to the plane ($X$-$Y$, $z= z_0(O_r)$). The direction of the electric field is opposite to the direction of EAS movement. The green arrow and green dots along the positive part of the $x$ axis show the order of variation of the coordinate $x_i$ of a circle of radius $R$ (first cycle), which allows us to calculate the number of seed EAS electrons sending electrons to the point $y_a(z)$; the pink arrow and pink dots along the negative part of the $x$ axis show the order of variation of the radius $R$ (second cycle), which allows us to calculate the sum of all electrons at the point $y_a(z)$; the red arrow and red dots along the positive part of the y axis show the order of variation of the coordinate of the point $y_a(z)$, in which the electron flux is calculated (third cycle); thick blue arrows show the distance $r^+$, $r^-$ in equation (s12) from the points ($x_i, y_i^+$), ($x_i, y_i^-$) to the point $y_a(z)$; symmetrical thin blue lines show the distances from the points ($x_i^-, y_i^+$), ($x_i^-, y_i^-$) to the point $y_a(z)$.

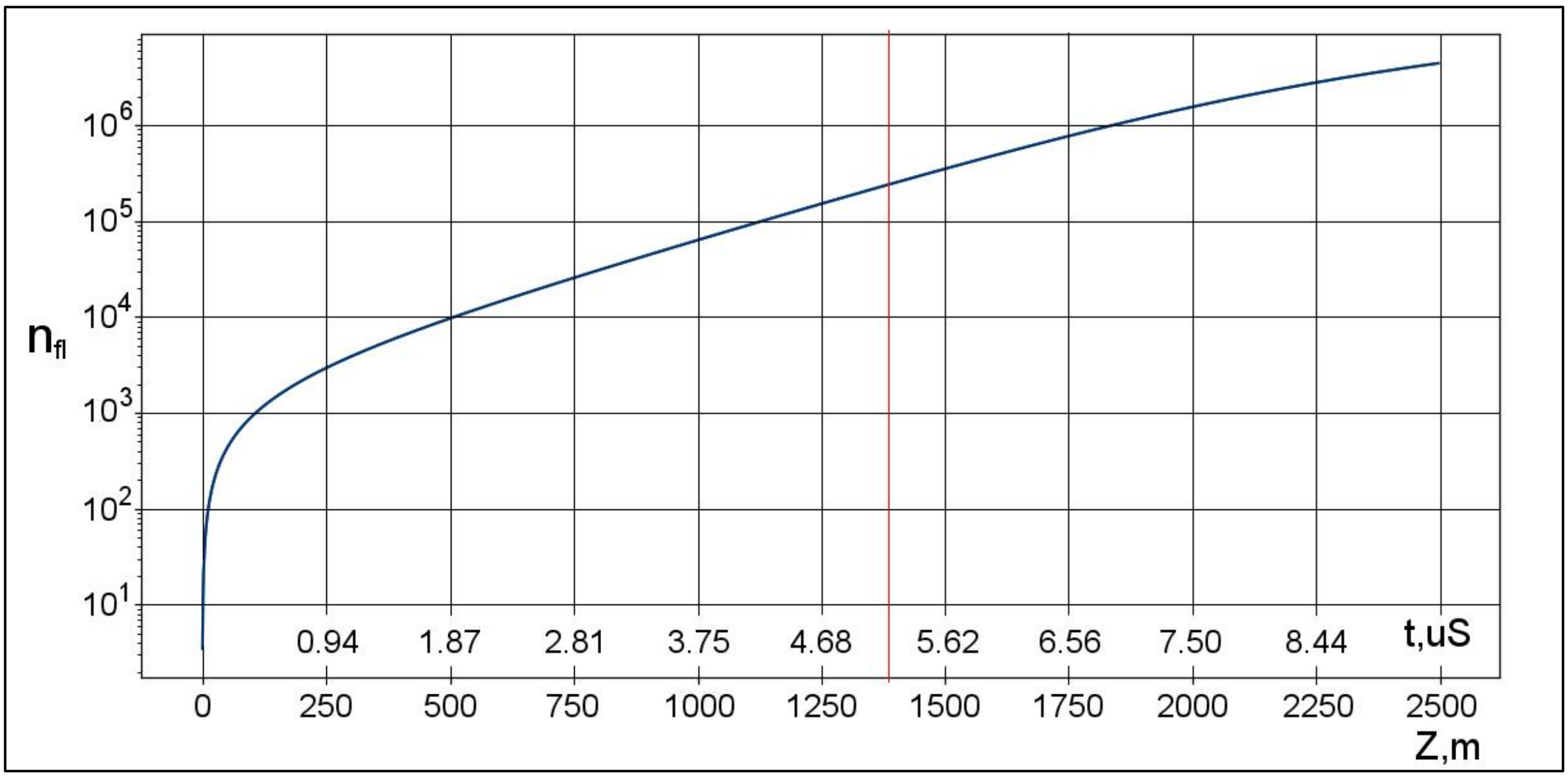

**Supplementary Figure S3.** Estimation of the number of streamer flashes $n_{fl}$ depending on the path *z(t)* that the EAS-RREA avalanche traverses inside the EE-volume. The calculation was carried out for a height of 13 km, $N_e^{EAS} = 10^6$, $R_M$=*328* m, $400\frac{kV}{m \cdot atm}\left(85\ \frac{kV}{m}\right)$, $\lambda_{RREA} = 277\ m$, the probability of initiation of air electrodes is $\mathrm{p} = 0.001$, the density of the number of air electrodes was considered constant and equal to $\rho_{\mathrm{E_{th}}} = 10^{-2}\mathrm{m}^{-3}$. The red vertical line shows the avalanche passage time 1385 m (5.4 µs from the beginning of the movement), which sets the volume of the EE-volume to ~1.0 км$^3$ ( $\mathrm{y_a(z)}$=500 м.)

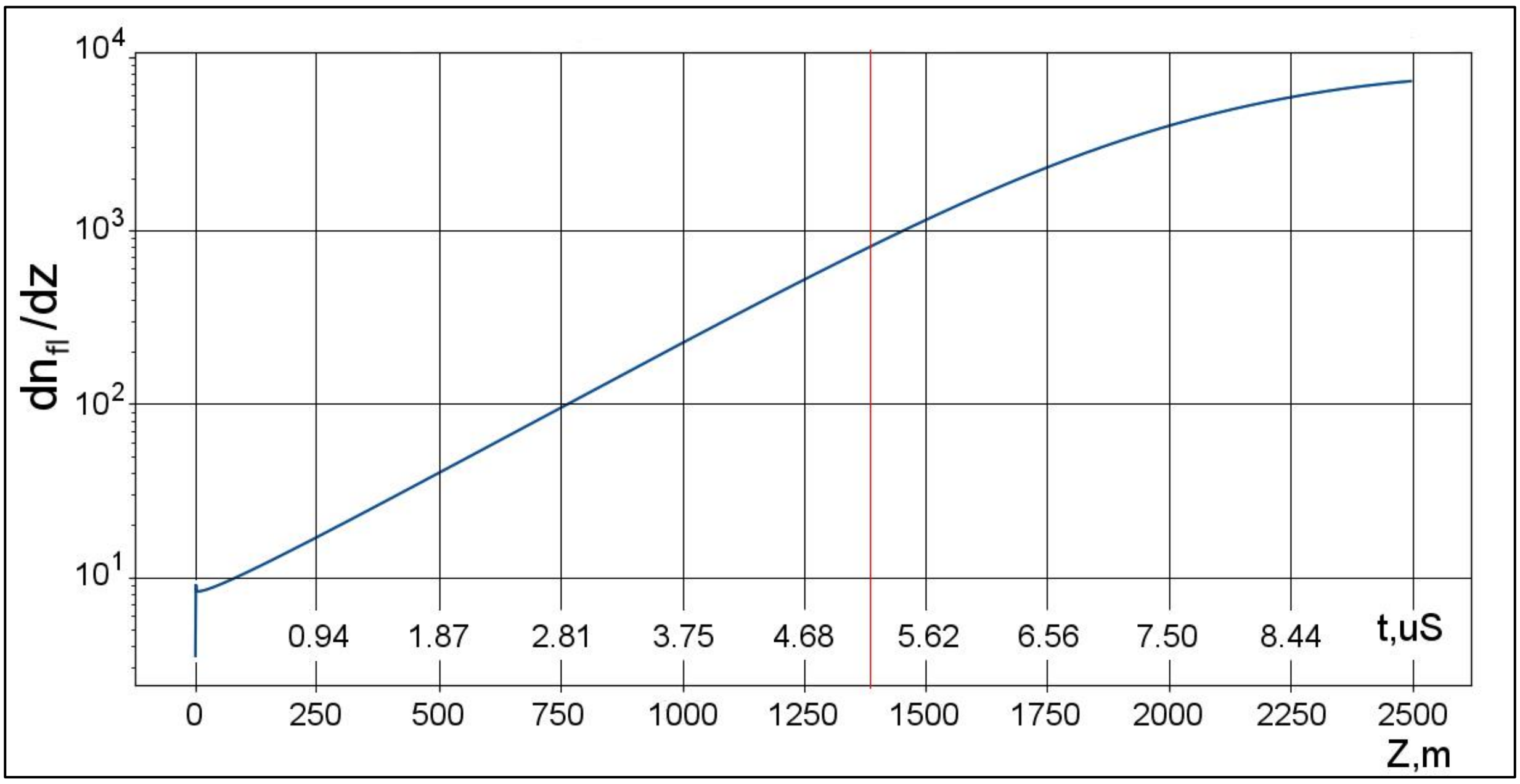

**Supplementary Figure S4.** The number of flashes $\mathrm{dn_{fl}/dz}$ in each transverse air layer 1 meter thick for the conditions of Figure S3: altitude is 13 km, $N_e^{EAS} = 10^6$, $R_M$=*328 m*, $400\frac{kV}{m \cdot atm}\left(85\ \frac{kV}{m}\right)$, $\lambda_{RREA} = 277\ m$, $\mathrm{p} = 0.001$, $\rho_{E_{th}} = 10^{-2}m^{-3}$.

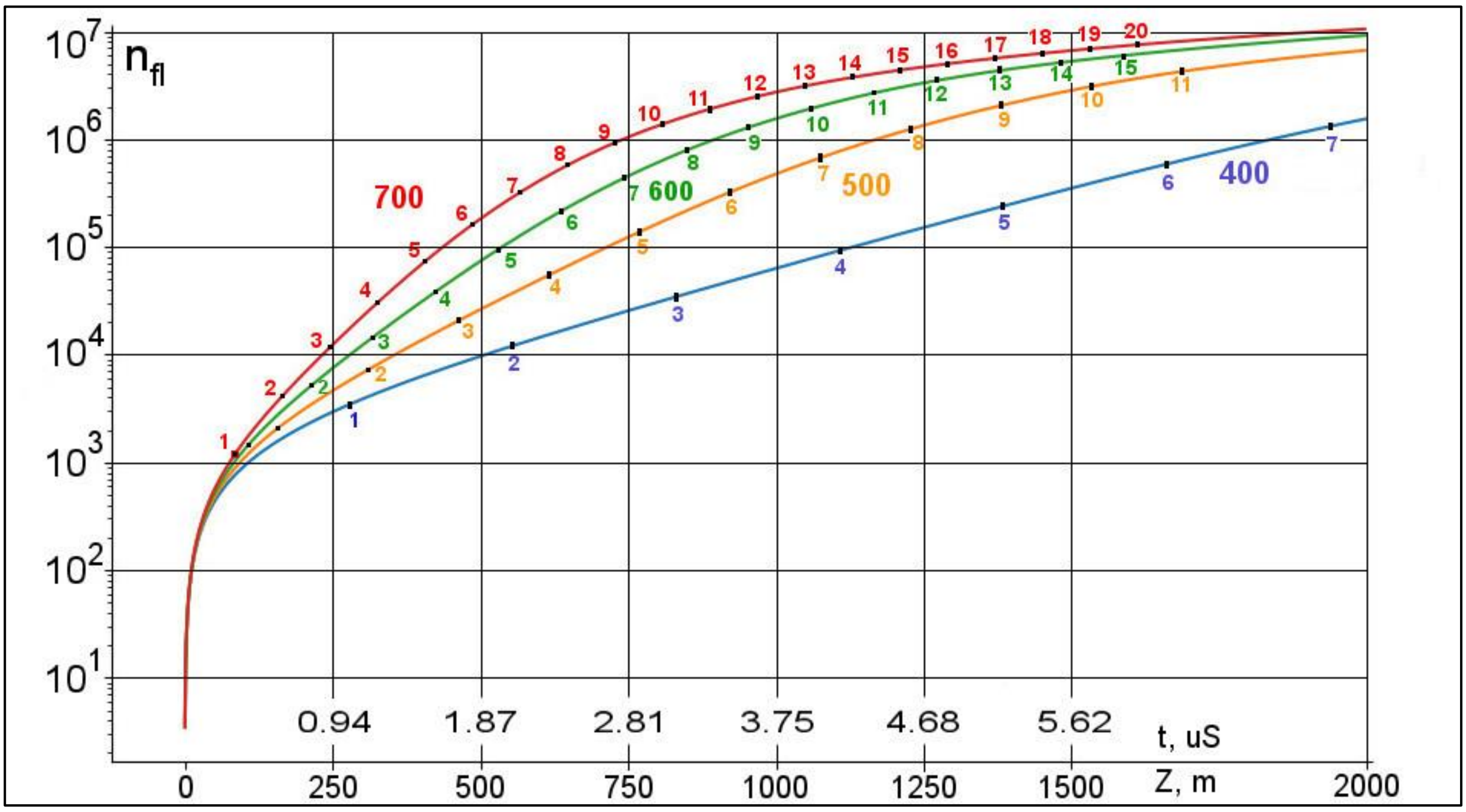

**Supplementary Figure S5.** Dependence of the number of streamer flashes $n_{fl}(z)$ on the electric field strength. Altitude 13 km, electron hit probability is $p = 0.001(\emptyset \approx 2\ cm), \rho_{E_{th}} = 10^{-2} m^{-3}$, $R_M$=328 m, $N_e^{EAS} = 10^6$. The electric field took four values: $400 \frac{kV}{m \cdot atm} \left(85\ \frac{kV}{m}\right)$, $\lambda_{RREA} = 277\ m$; — blue line; $500 \frac{kV}{m \cdot atm} \left(106 \frac{kV}{m}\right)$, $\lambda_{RREA} = 153\ m$ — yellow line; $600 \frac{kV}{m \cdot atm} \left(127.5\ \frac{kV}{m}\right)$, $\lambda_{RREA} = 106\ m$ — green line; $700 \frac{kV}{m \cdot atm} \left(149\ \frac{kV}{m}\right)$, $\lambda_{RREA} = 81\ m$ — red line. The numbers near the lines indicate the step number of the avalanche $\lambda_{RREA}$.

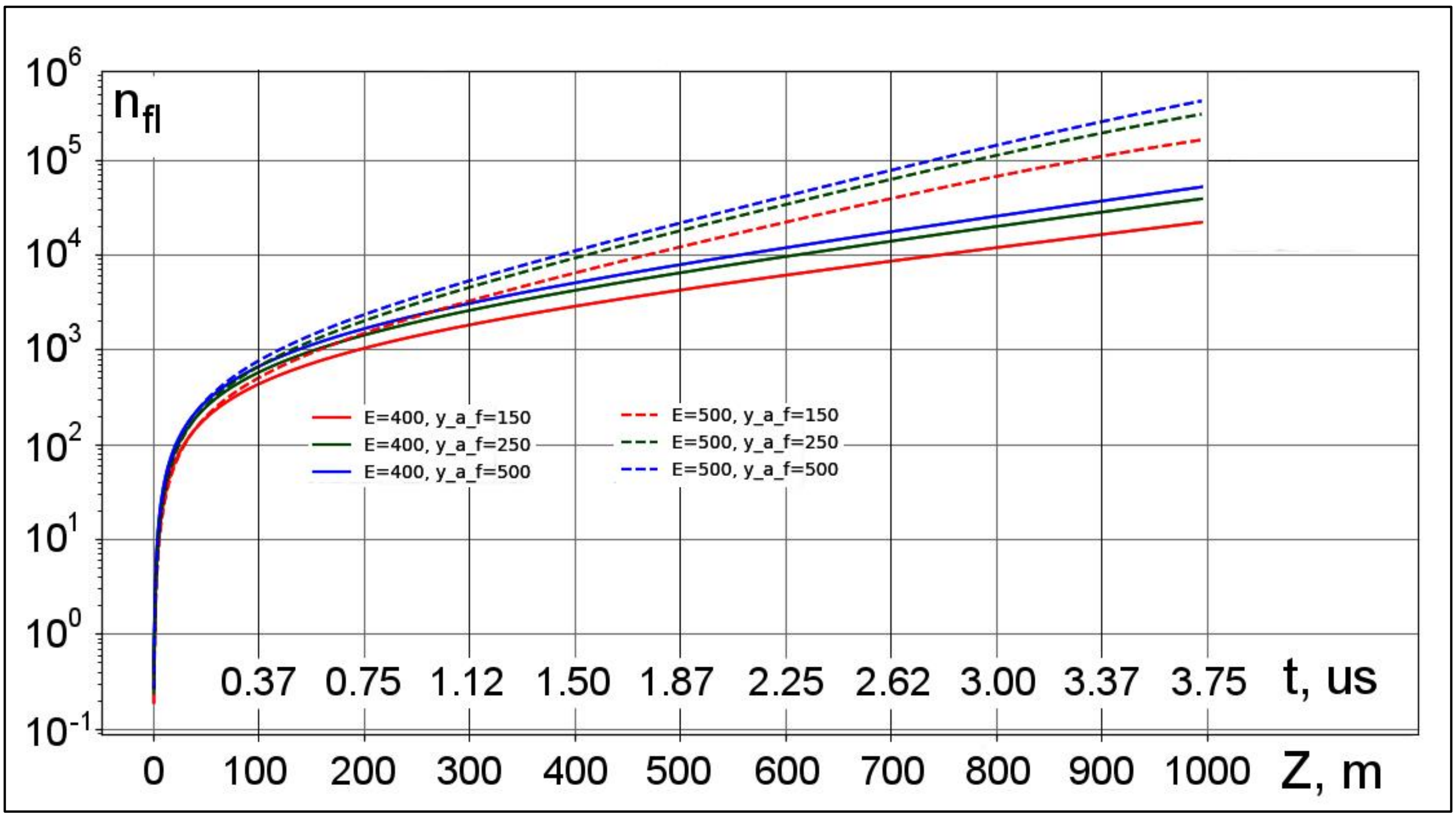

**Supplementary Figure S6.** Effect on the number of streamer flashes $n_{fl}(z)$ of volumes of the EE-volume of a thundercloud (at an altitude of 13 km), which are defined by the radii of the EE-volume: $y_{aF} = 500\ m$ (blue line), $y_{aF} = 250\ m$ (green line) и $y_{aF} = 150\ m$ (red line), for electric fields $400 \frac{kV}{m \cdot atm} \left(85\ \frac{kV}{m}\right)$, $\lambda_{RREA} = 277\ m$ — solid lines, $500 \frac{kV}{m \cdot atm} \left(106 \frac{kV}{m}\right)$, $\lambda_{RREA} = 153\ m$) — dashed lines. The electron hit probability is $p = 0.001(\emptyset \approx 2\ cm), \rho_{E_{th}} = 10^{-2} m^{-3}$, $R_M$=328 m, $N_e^{EAS} = 10^6$.

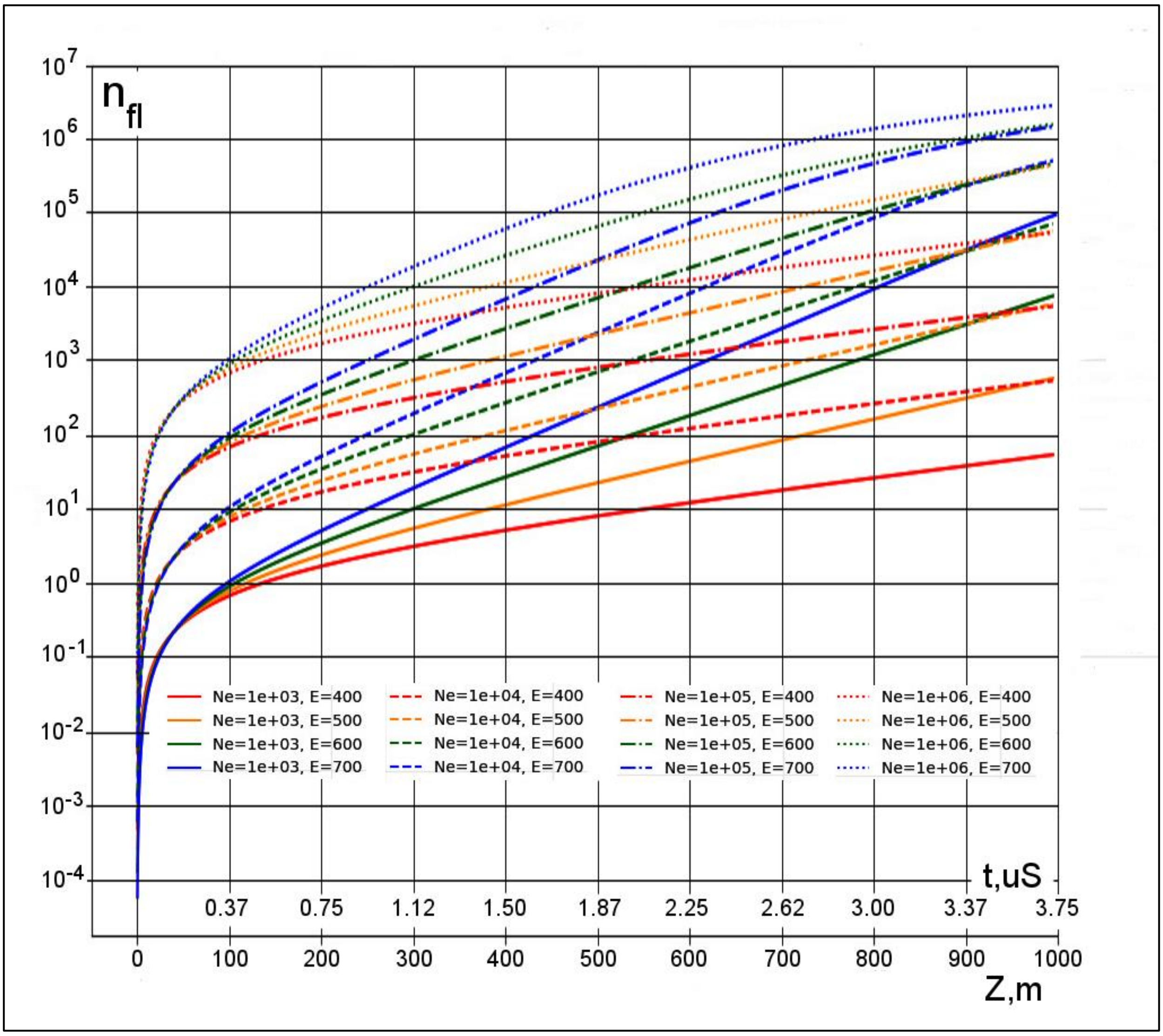

**Supplementary Figure S7.** The number of streamer flashes $n_{fl}(z)$ at an altitude of 13 km depending on the number of EAS seed electrons $N_e^{EAS} = 10^3 (solid\ lines)$; $10^4$ (dashed lines); $10^5$ (dashed-dotted lines), $10^6$ (dotted lines). For each of these $N_e^{EAS}$ values, the calculation was performed for four electric fields: $400 \frac{kV}{m \cdot atm}\left(85 \frac{kV}{m}\right)$, $\lambda_{RREA} = 277\ m$ — red line; $500 \frac{kV}{m \cdot atm}\left(106 \frac{kV}{m}\right)$, $\lambda_{RREA} = 153\ m$ — yellow line; $600 \frac{kV}{m \cdot atm}\left(127.5 \frac{kV}{m}\right)$, $\lambda_{RREA} = 106\ m$ — green line; $700 \frac{kV}{m \cdot atm}\left(149 \frac{kV}{m}\right)$, $\lambda_{RREA} = 81\ m$ — blue line. Height is 13 km, probability of electron hit is $p = 0.001 (\emptyset \approx 2\ cm)$, $\rho_{E_{th}} = 10^{-2} m^{-3}$, $R_M$=328 m.

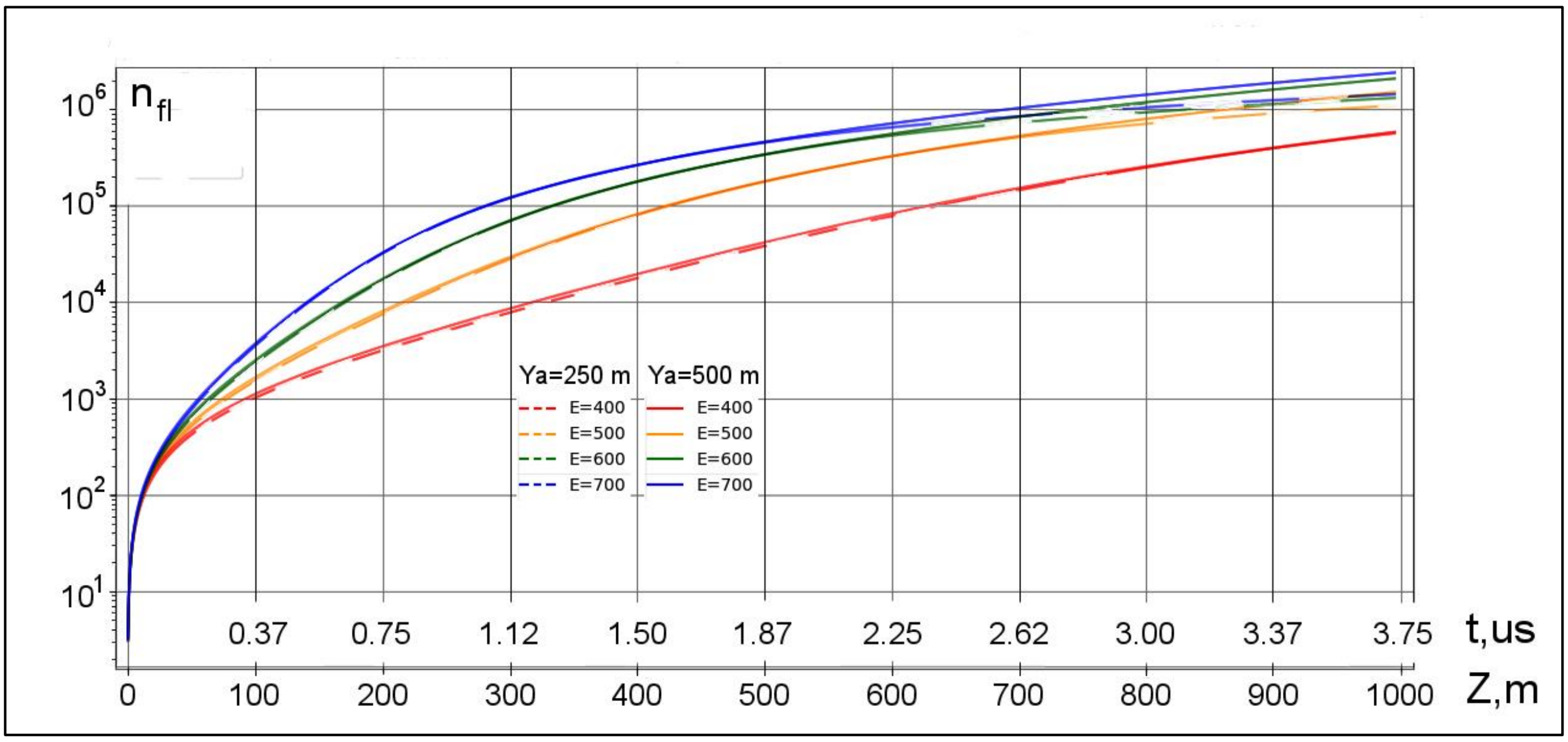

**Supplementary Figure S8.** The influence of the sizes of the EE-volume of a thundercloud (height 6 km, $N_e^{EAS} = 10^6$) on the number of streamer flashes $n_{fl}$(z). The dimensions of the EE-volume are determined by the radii $y_{aF} = 250\, m\ (dashed\ lines), y_{aF} = 500\, m\ (solid\ lines)$: $400\frac{kV}{m \cdot atm}\left(196\ \frac{kV}{m}\right),\ \lambda_{RREA} = 120\, m$ — red lines; $500\frac{kV}{m \cdot atm}\left(244\frac{kV}{m}\right), \lambda_{RREA} = 66\, m$ — yellow lines; $600\frac{kV}{m \cdot atm}\left(294\ \frac{kV}{m}\right), \lambda_{RREA} = 46\, m$ — green lines; $700\frac{kV}{m \cdot atm}\left(343\ \frac{kV}{m}\right), \lambda_{RREA} = 35\, m$ — blue lines. The electron hit probability is $p = 0.001(\emptyset \approx 2\ cm), \rho_{E_{th}} = 10^{-2} m^{-3}$, $R_M$=161 .

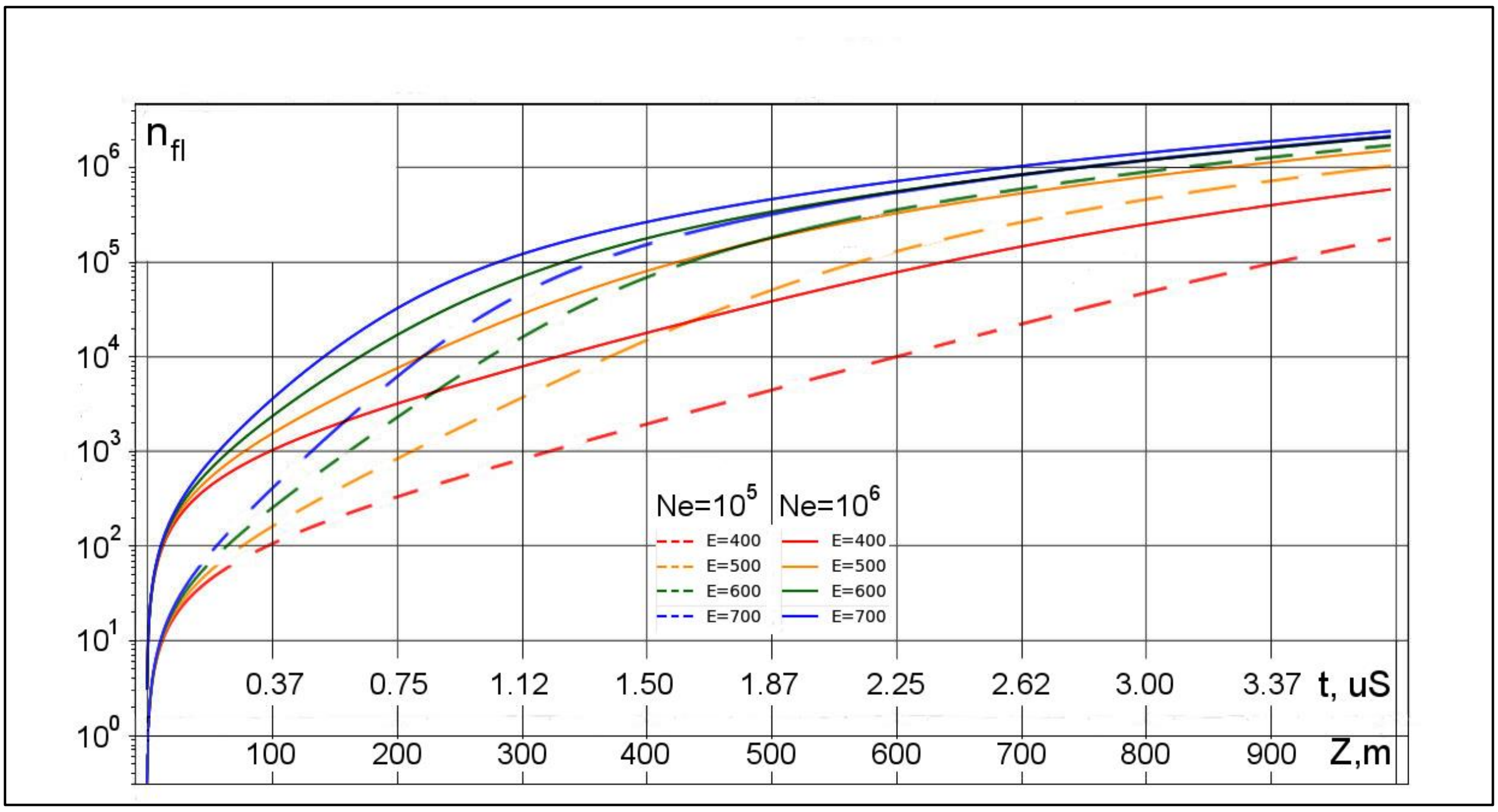

**Supplementary Figure S9.** The number of streamer flashes $n_{fl}$(z) at an altitude of 6 km depending on the number of EAS seed electrons ($N_e^{EAS} = 10^5 (dashed\ lines),\ 10^6 (solid\ lines)$. For each of these values $N_e^{EAS}$ the calculation was performed for four electric fields: $400\frac{kV}{m \cdot atm}\left(196\ \frac{kV}{m}\right),\ \lambda_{RREA} = 120\, m$ — red lines; $500\frac{kV}{m \cdot atm}\left(244\frac{kV}{m}\right),\ \lambda_{RREA} = 66\, m$ — yellow lines; $600\frac{kV}{m \cdot atm}\left(294\ \frac{kV}{m}\right),\ \lambda_{RREA} = 46\, m$ — green lines; $700\frac{kV}{m \cdot atm}\left(343\ \frac{kV}{m}\right),\ \lambda_{RREA} = 35\, m$ — blue lines. The probability of electron hit is $p = 0.001(\emptyset \approx 2\ cm), \rho_{E_{th}} = 10^{-2} m^{-3}$, $R_M$=161 m.